\pdfoutput=1
\documentclass[aps,prl,reprint,groupedaddress,floatfix]{revtex4-1}

\usepackage{graphicx}
\usepackage{dcolumn}
\usepackage{bm}
\usepackage{physics}
\usepackage{amssymb}
\usepackage{natbib}

\begin{document}
	
	\title{Laminar Flow and Backrind Formation in Molding of Viscoelastic Silicone Rubber}
	
	
	\author{Louis A. Bloomfield}
	\email{lab3e@virginia.edu}
	\affiliation{Department of Physics, University of Virginia, Charlottesville, VA 22904}
	
	
	\date{\today}
	
	\begin{abstract}
		When a thermoset polymer is cured at elevated temperature in a closed mold, thermal expansion can produce flaws in the finished product. Those flaws occur when rising internal pressure pushes the mold open and cured polymer flows out through gaps at the parting lines. Known as backrind, such defects are particularly common in compression molding, where the increasing pressure of a trapped, incompressible polymer can overwhelm the clamping pressure on the mold and expel polymer from the mold pocket. If that ejected material has already cured, it leaves behind structural damage and consequently a flaw in the finished product.

		Backrind usually appears as a ragged seam line near the gap where cured polymer exited the mold. Its appearance is typically irregular and fragmented, suggesting no particular pattern or uniformity to the process that produced it. In such cases, the cured polymer acts predominantly as a viscoelastic solid as it is driven toward and through the parting line. The backrind's ragged character results from tearing and fragmentation of that solid.
		
		It is possible, however, for the cured polymer to act predominantly as a viscoelastic liquid as it flows toward and through the parting line. Since the Reynolds number is low, the flow is laminar and the backrind bears witness to that laminar flow. More specifically, the backrind's observed shaped corresponds to isochronous contours in the laminar flow toward the parting line, contours that can be predicted using computational fluid dynamics.
	\end{abstract}
	
	\pacs{}
	
	\maketitle
	
	\section{Introduction}
	Many commercial products are made from thermoset polymers that have been molded into shape at elevated temperatures. The increasing temperatures inherent in this curing process present a fundamental challenge to manufacturing. Whereas thermoplastic polymers are introduced into molds as hot liquids and undergo thermal contraction as they solidify by cooling, thermosets enter molds as cool liquids and undergo thermal expansion as they solidify by heating. Starting with a full mold and having its contents expand can lead to trouble.
	
	Since liquid polymers are nearly incompressible, the pressure of a trapped polymer's frustrated thermal expansion can overwhelm the mold's clamping pressure and cause the mold to open slightly. Polymer can then leak out through resulting gaps at the parting lines. Because of its proximity to the hot mold surface, the ejected polymer may have already cured and, if so, the product will develop a flaw known as backrind.

	Backrind is normally minimized by metering and shaping the uncured ``preps'' placed in molds, decreasing the curing temperatures, preheating the preps, chemically retarding the cure, adding momentary openings or ``bumps'' to the molding sequence and, as a last resort, adding features to the mold that direct the backrind to where it doesn't matter or can be trimmed away.\cite{stritzke2012custom}
	
	When backrind does occur, it indicates that three things have occurred: that polymer near the parting line cured before the mold's contents reached thermal equilibrium, that continued thermal expansion produced an opening pressure that exceeded the clamping pressure, thereby opening the mold slightly, and that the outward pressure gradient defeated the inward viscoelastic stress and propelled cured polymer toward and through gaps at the parting lines.
	
	Different polymers experience different viscoelastic stresses during backrind formation. In a thermosetting polymer that cures to a strong, rigid plastic, the inward viscoelastic stresses in the cured plastic near the parting lines are predominantly static elastic in character and capable of withstanding intense pressure gradients. Since the cured plastic barely moves when gaps open at the parting lines, there is almost no ejected material. Hard thermoset plastics are thus unlikely to develop significant backrind.
	
	In a thermosetting polymer that cures to a pliable rubber, however, backrind is a serious possibility. The inward viscoelastic stresses in the cured rubber near the parting lines are again predominantly static elastic, but they are not strong enough to withstand intense pressure gradients. When gaps appear at the parting lines, enormous pressure gradients propel nearby rubber toward and through those gaps. Some of the rubber may distort beyond its elastic limits, tearing or deforming permanently, so damaged and distressed rubber is likely to be found in the seam line region. Pliable thermoset rubbers are thus susceptible to backrind.
	
	To observe a third type of backrind, it is necessary to consider another class of thermosetting polymers, polymers that cure to viscoelastic rubbers in which dynamic viscoelastic stresses are overwhelmingly stronger than static elastic stresses, except at the slowest strain rates or during the most prolonged strains. Only in such rubbers can the inward viscoelastic stresses in cured rubber near the parting line be predominantly dynamic viscoelastic in character. This class of thermosetting polymers can be found among the viscoelastic silicone rubbers.
	
	\section{Viscoelastic Silicone Rubber}
	Viscoelastic silicone rubbers (VSRs) are silicone elastomers in which the silicone polymer chains are networked together by both permanent and temporary crosslinks.\cite{bloomfield2018borosilicones} A VSR's dynamic viscoelastic stresses are due primarily to its temporary crosslinks and they can be quite strong. Its static elastic stresses are due only to its permanent crosslinks and they can be quite weak or even zero.
	
	For a VSR to support static elastic stresses, its network of permanent crosslinks must exceed the gelation threshold.\cite{flory1941a,flory1941b,stockmayer1944} The VSR is then a network solid---a gigantic macromolecule in which a single covalently-bonded network extends throughout the entire material. Below the gelation threshold, the VSR is a network liquid---a gigantic macromolecule in which a single covalently-bonded network extends throughout the material, but with covalent bonds that detach and re-attach frequently so that the network can evolve in topology and geometry. An uncured VSR is a network liquid that cannot support static elastic stresses, a cured VSR is a network solid that can support elastic static stresses, and curing a VSR amounts to forming enough new permanent crosslinks to cross the gelation threshold between network liquid and network solid. 
	
	VSRs can be divided loosely into two types, firm and soft, based on the ratio of static elastic stress to dynamic viscoelastic stress after a sudden change in strain. In a firm VSR, that ratio is relatively large and the substantial static elastic stress can be differentiated quickly from the dynamic viscoelastic stress. In a soft VSR, that ratio is relatively small and the meager static elastic stress cannot be differentiated quickly from the dynamic viscoelastic stress. In fact, it can take considerable time to determine whether or not a soft VSR's static elastic stress is actually greater than zero. 
	
	This article henceforth focuses only on \textit{soft} VSRs (SVSRs). That narrowed focus has to do with molding and backrind formation. Only in cured SVSRs are viscoelastic stresses so overwhelmingly dynamic viscoelastic in character that static elastic stresses can be ignored at all but the slowest strain rates or during the most prolonged strains. When gaps appear at parting lines and pressure gradients propel cured SVSR toward and through those gaps, static elastic stresses are negligible and the solid material's motion is indistinguishable from viscous flow in a liquid. 
	
\begin{figure}
	\includegraphics{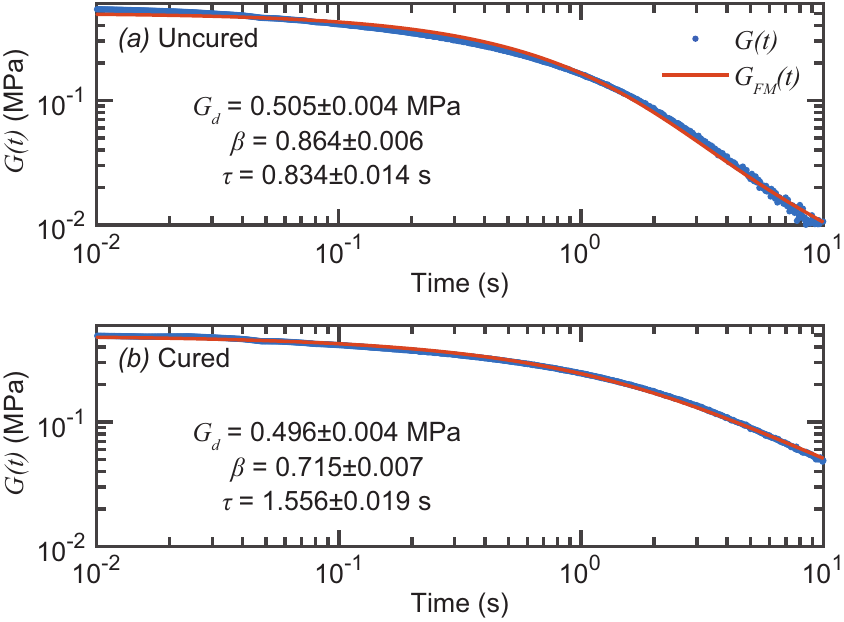}
	\caption{Stress relaxation modulus $G(t)$ of an SVSR in its (a) uncured and (b) cured states, fit by the stress relaxation modulus $G_{\textsc{fm}}(t)$ of the Fractional Maxwell model. The dynamic modulus G$_{d}$, the characteristic time $\tau$, and the fractional order $\beta$ are the fit's three parameters and $1\ge\beta> 0$. Only the first 10 seconds of each measurement are shown and considered in the fitting process. \label{fig:figShortCompressionMeasurements}}
	\end{figure}

	Treating a solid as a liquid is so self-contradictory that it demands further justification. Toward that end, we compare a cured (solid) SVSR and its uncured (liquid) form and observe that, on short timescales or at high frequencies, both behave as elastic fluids.\cite{green1946,shaw2012,bloomfield2018borosilicones} Figure \ref{fig:figShortCompressionMeasurements} shows the stress relaxation modulus $G(t)$ for the SVSR in its (a) uncured and (b) cured forms during the first 10 seconds after a sudden step in compression. Although the measured curves differ quantitatively, they are similar in shape and both are consistent with $G(t)\rightarrow 0$ as $t\rightarrow \infty$. There is as yet no evidence of solid behavior in either form of the SVSR, uncured or cured.
	
	As shown in \citet{bloomfield2018borosilicones}, the behavior of a liquid VSR can be described by the Fractional Maxwell viscoelastic model. The stress relaxation modulus $G_{\textsc{fm}}(t)$ of that model is
	\begin{equation}
	G_{\textsc{fm}}(t) = G_{d}E_{\beta}\left(-(t/\tau)^{\beta}\right),\label{eq:FMaxwellStressRelaxation}
	\end{equation}
	where 
	$E_{\beta}(z)$ is the Mittag-Leffler function,
	\begin{equation}
	E_{\beta}(z) = \sum_{k=0}^{\infty}\frac{z^{k}}{\Gamma(\beta k + 1)},
	\end{equation}
	and the dynamic modulus $G_{d}$, the characteristic time $\tau$, and the fractional order $\beta$ are the model's three parameters and $1\ge\beta> 0$. Fig. \ref{fig:figShortCompressionMeasurements} includes values of those parameters, obtained by fitting Eq. \ref{eq:FMaxwellStressRelaxation} to the measured $G(t)$.
	
	The fits are quite good and indicate that both materials behave in accordance with the Fractional Maxwell model for at least the first 10 seconds following a compression step. As expected of a liquid, $G_{\textsc{fm}}\rightarrow 0$ as $t\rightarrow\infty$. There is no sign yet that cured SVSR can support static elastic stress.
	
\begin{figure}
		\includegraphics{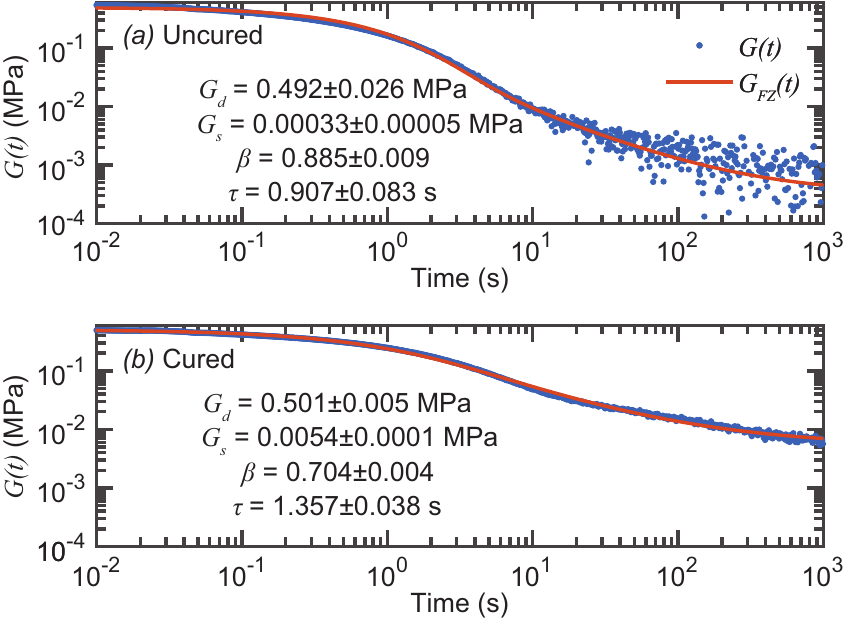}
		\caption{Stress relaxation modulus $G(t)$ of an SVSR in its (a) uncured and (b) cured states, fit by the stress relaxation modulus $G_{\textsc{zm}}(t)$ of the Fractional Zener model. The dynamic modulus G$_{d}$, the static modulus $G_{s}$, the characteristic time $\tau$, and the fractional order $\beta$ are the fit's four parameters and $1\ge\beta> 0$. All 1000 seconds of each measurement are shown and considered in the fitting process. \label{fig:figCompressionMeasurements}}
	\end{figure}

	It is only on long timescales or at low frequencies that liquid and solid VSRs become distinguishable. Figure \ref{fig:figCompressionMeasurements} shows the same $G(t)$ measurements as in Fig. \ref{fig:figShortCompressionMeasurements}, except during a longer period: 1000 seconds. In this extended time frame, it can be seen that both $G(t)$ curves bend upward at later times and it is unclear whether they go to zero or remain finite as $t\rightarrow\infty$.

	The Fractional Maxwell viscoelastic model describes only liquids, however, it can be extended to describe solids by adding a static elastic element. The result is the Fraction Zener viscoelastic model.\cite{bloomfield2018borosilicones} The stress relaxation modulus $G_{\textsc{fz}}(t)$ of that model is
	
	\begin{equation}
	G_{\textsc{fz}}(t) = G_{s}+G_{d}E_{\beta}\left(-(t/\tau)^{\beta}\right),\label{eq:FZenerStressRelaxation}
	\end{equation}
	where the dynamic modulus $G_{d}$, the static modulus $G_{s}$, the characteristic time $\tau$, and the fractional order $\beta$ are the model's four parameters and $1\ge\beta> 0$. Fig. \ref{fig:figCompressionMeasurements} includes values of those parameters, obtained by fitting Eq. \ref{eq:FZenerStressRelaxation} to the measured $G(t)$.
	
	The fits are again quite good, but this time they indicate that both materials behave in accordance with the Fractional Zener model for at least the first 1000 seconds following a compression step. The static modulus $G_{s}$ of cured SVSR is unambiguously greater than zero, consistent with this material being a network solid with an equilibrium shape to which it returns in the absence of external influences and the ability to support static elastic stress. Nonetheless, those solid properties are relatively weak and therefore difficult to observe on molding timescales.
	
	The static modulus $G_{s}$ of uncured SVSR is barely greater than zero. It is definitely a network liquid, being below the gelation threshold, and it has no equilibrium shape. Nonetheless, it does not flow under its own weight. Because it contains fumed silica particles, it is likely a Bingham plastic liquid, in which the fumed silica particles gradually forming weak solid structures in static conditions and allow it to retain weak static elastic stresses during long measurements.

	\section{Backrind Formation in SVSR}

	When rising internal pressure opens its mold, cured SVSR flows like a liquid toward and through gaps in the parting lines. Its kinematic viscosity is about 100 m$^{2}$/s, its velocity is about $10^{-3}$ m/s, and the characteristic length of its environment is about $10^{-3}$ m, so the flow has a Reynolds number of about $10^{-8}$. The cured SVSR thus experiences dissipative laminar flow toward and through the gaps.

	Even though the cured SVSR's permanently crosslinked polymer network is too weak to influence its flow, that network can experience damage if strained beyond its elastic limits. Excessive strain can thus alter the SVSR's equilibrium shape. The network is likely to remain intact during the SVSR's approach to the gap because the flow in that relatively open region necks down gradually, with only modest shear and extension. In the narrow gap itself, however, there are enormous velocity gradients across the flow associated with stationary fluid at the walls and fast moving fluid at the midpoint. In this high-shear flow, the network is easily torn to shreds and the cured SVSR's equilibrium shape forgotten. 
	
	Therefore, when thermal expansion pushes a mold open and the resulting pressure gradient propels cured SVSR toward a gap in the parting line, that SVSR divides into two portions. The portion that enters the gap loses most of its elastic network and its equilibrium shape, and it exhibits little attachment to the product when the mold is opened. The portion of SVSR that does not enter the gap retains its elastic network and slowly returns to its equilibrium shape within the product when the mold is opened. The product gradually develops an indentation or groove, in which the missing volume corresponds to the cured SVSR that was extruded through the gap and did not return. SVSRs are thus quite susceptible to backrind and that backrind has a structure that is directly associated with laminar flow.
	
	The groove's skin is the forwardmost surface of the returning portion, the layer of cured SVSR that just barely avoided entering the parting-line gap and losing its elastic component. That layer had just arrived at the gap at the moment the extrusion process ceased. The points on the groove's skin thus have something in common: they all underwent laminar flow toward the gap and arrived at that gap simultaneously, at the final moment of flow. The groove's skin is thus an isochron of the laminar flow: a surface whose points took equal times to reach the gap's entrance.
	
	Since the SVSR undergoes dissipative laminar flow and its elastic network has negligible effect on that flow, this situation can be modeled easily with Computational Fluid Dynamics (CFD). Moreover, the detailed geometry of the mold pocket and parting line gap has little effect on the results. All that really matters is that a viscous fluid in an open channel flows toward and through a narrow slot under the influence of a pressure difference and that a certain volume of SVSR is ejected per unit length of the slot. As long as the open channel is much wider than the slot, their widths, along with the specific pressure difference and SVSR viscosity, hardly matter at all. Furthermore, the results for transient flow and steady state flow do not differ significantly.
	
\begin{figure}
		\includegraphics[width=0.7\columnwidth]{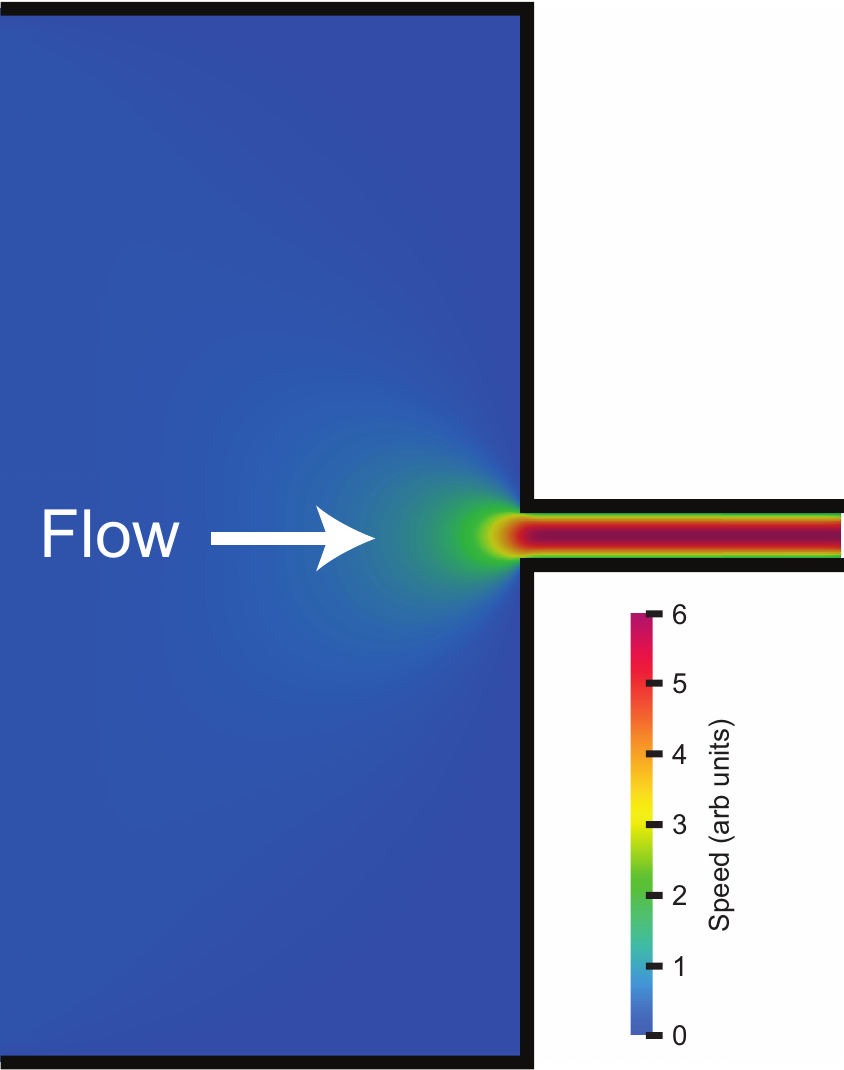}
		\caption{The speed field (i.e., velocity magnitude field) for a viscous fluid flowing from a wide channel to a narrow slot under the influence of a pressure difference between entrance and exit. Produced by the paraFoam visualizer from the SimpleFoam solution.\label{fig:figSpeed}}
	\end{figure}

\begin{figure}
		\includegraphics[width=0.7\columnwidth]{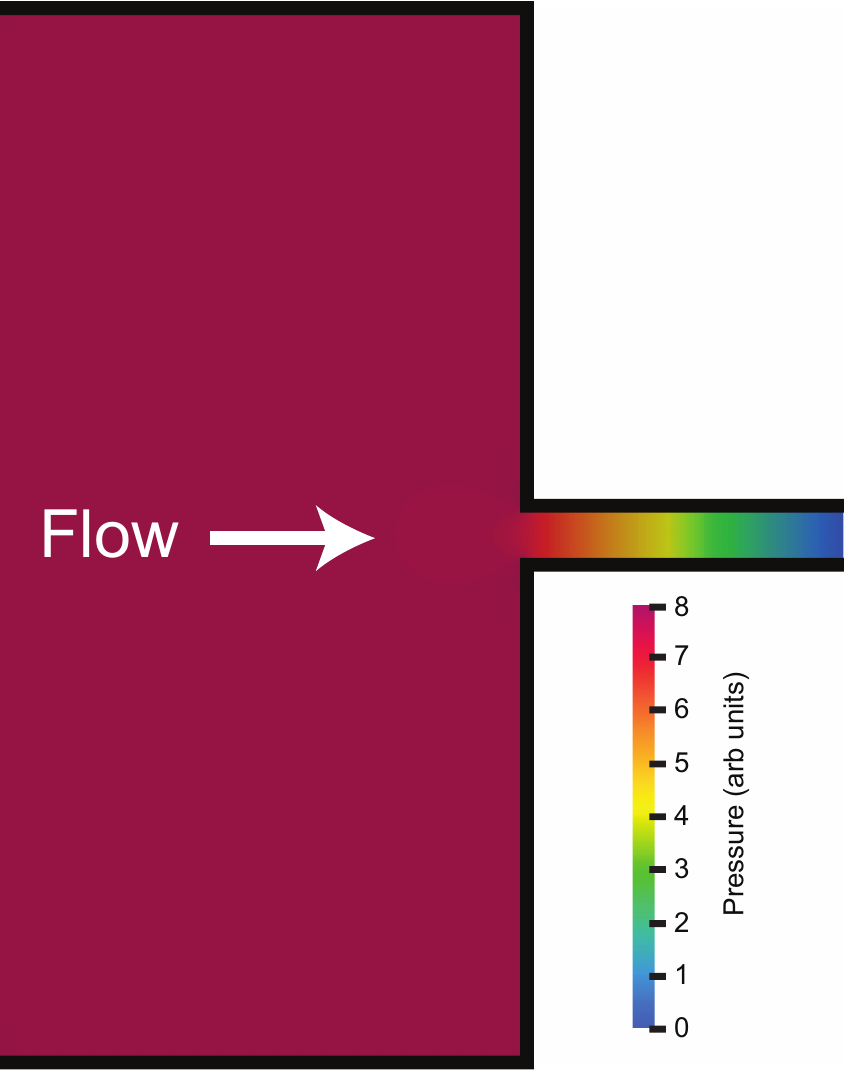}
		\caption{The pressure field for a viscous fluid flowing from a wide channel to a narrow slot under the influence of a pressure difference between entrance and exit. Produced by the paraFoam visualizer from the SimpleFoam solution.\label{fig:figPressure}}
	\end{figure}

\begin{figure}
		\includegraphics[width=0.7\columnwidth]{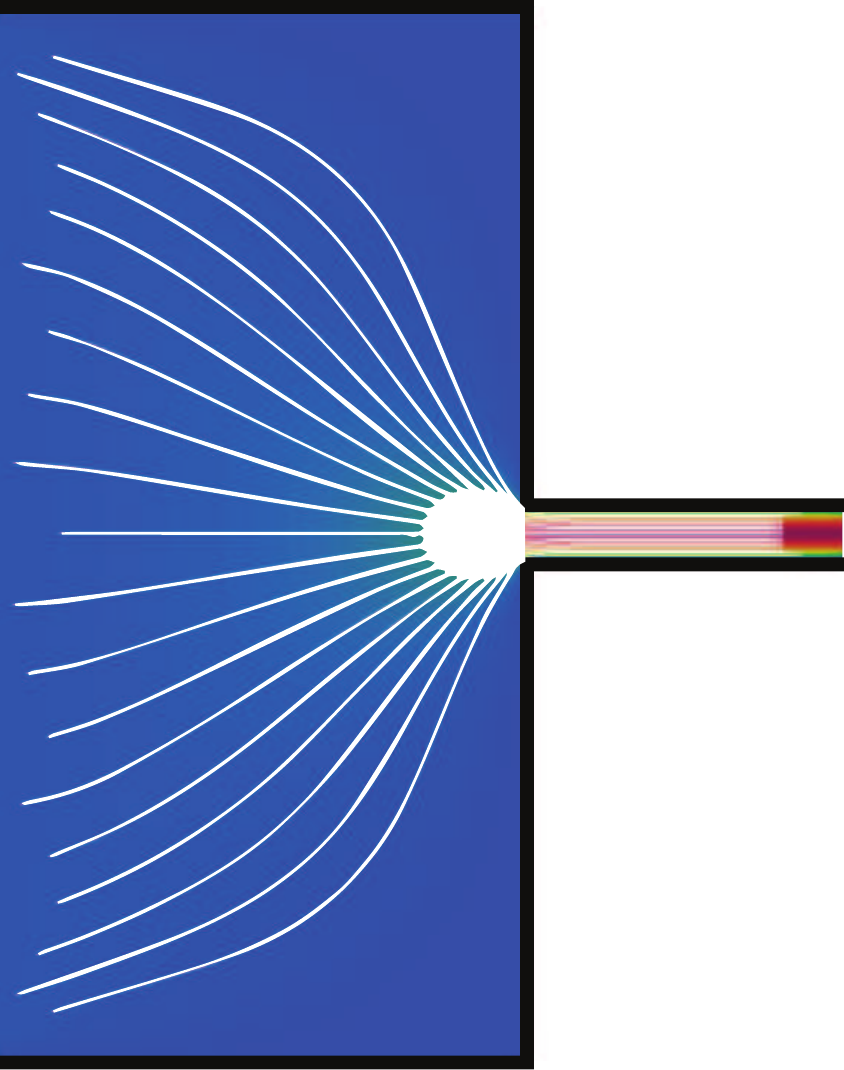}
		\caption{Streamlines (white) for a viscous fluid flowing from a wide channel to a narrow channel/slot under the influence of a pressure difference between entrance and exit, superposed on the flow's speed field. Produced by the paraFoam post-processor and visualizer from the SimpleFoam solution.\label{fig:figStreamlines}}
	\end{figure}

	The OpenFoam CFD package was used to model viscous flow from a wide channel to a narrow slot under the influence of a pressure drop from entrance and exit. A steady-state incompressible laminar solution to that model was obtained using the SimpleFoam solver. Because the model extended infinitely in the third dimension, it was effectively two-dimensional and its results were two-dimensional cross sections. Figure \ref{fig:figSpeed} and \ref{fig:figPressure} show the flow's velocity magnitude field and pressure field respectively, while Fig. \ref{fig:figStreamlines} shows the flow's streamlines. 
	
	The three figures tell a consistent story. As the fluid in the wide channel approaches the narrow slot, its streamlines neck together, its speeds increase, and its pressures drop slightly. There is little dissipation in the wide channel, so only a small drop in pressure is needed to produce the speed increase.
	
	Inside the narrow slot, however, viscous interactions with the nearby walls produce the severe velocity gradients and rapid dissipation of Poiseuille flow. The fluid's pressure plummets in the downstream direction because a steep pressure gradient is needed to keep it moving downstream at constant velocity. Its speed reaching a maximum value in the middle of the slot, so most of the fluid volume in this flow travels along the middle streamlines of Fig. \ref{fig:figStreamlines}.
	
\begin{figure}
		\includegraphics[width=0.7\columnwidth]{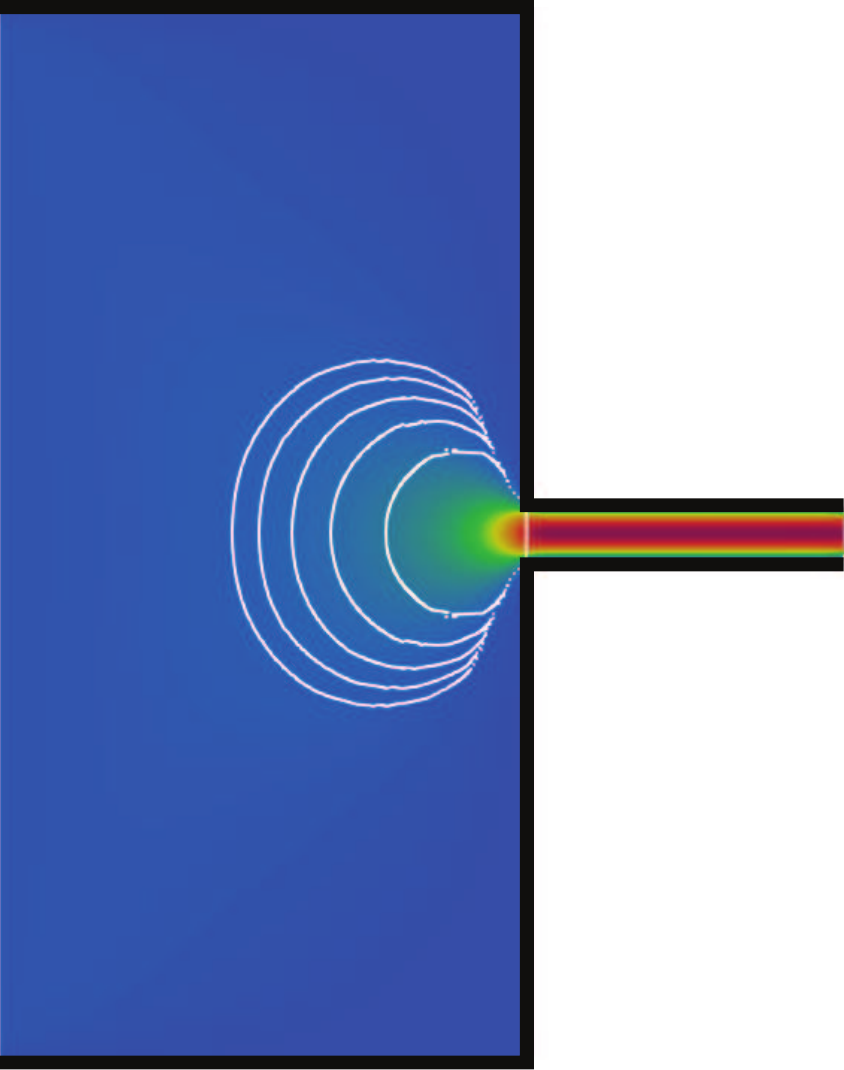}
		\caption{Caption. \label{fig:figContours}}
	\end{figure}

	Having solved the CFD model to obtain the fluid's complete velocity field, it is possible to trace the motion of any portion of the fluid backward or forward in time. Specifically, one can start with the sheet of fluid at the entrance to the slot at one time and calculate where that same sheet was at an earlier time. Earlier-time sheets obtained by this procedure (Fig. \ref{fig:figContours}) are isochrons---surfaces of fluid that are separated from the slot's entrance by specific amounts of time. They are surfaces of fluid that will flow to reach the slot's entrance in those amounts of time.

	Recall that the skin of an SVSR's backrind groove consists of the fluid that reached the entrance to the parting-line gap at the moment flow ceased. When allowed to return to its equilibrium shape, that fluid returns to where it cured before the flow commenced and becomes the skin of the backrind groove.
	
	Thus the skin of the backrind groove is one of the isochrons. It is the isochron that is separated from the gap entrance by duration of the flow. It is also the isochron that encompasses the volume of fluid per unit length extruded into the slot during the backrind formation. The more volume per unit length that exits the mold through the parting-line gap, the earlier the isochron corresponding to the backrind groove's skin.
	
\begin{figure}
		\includegraphics[width=0.5\columnwidth]{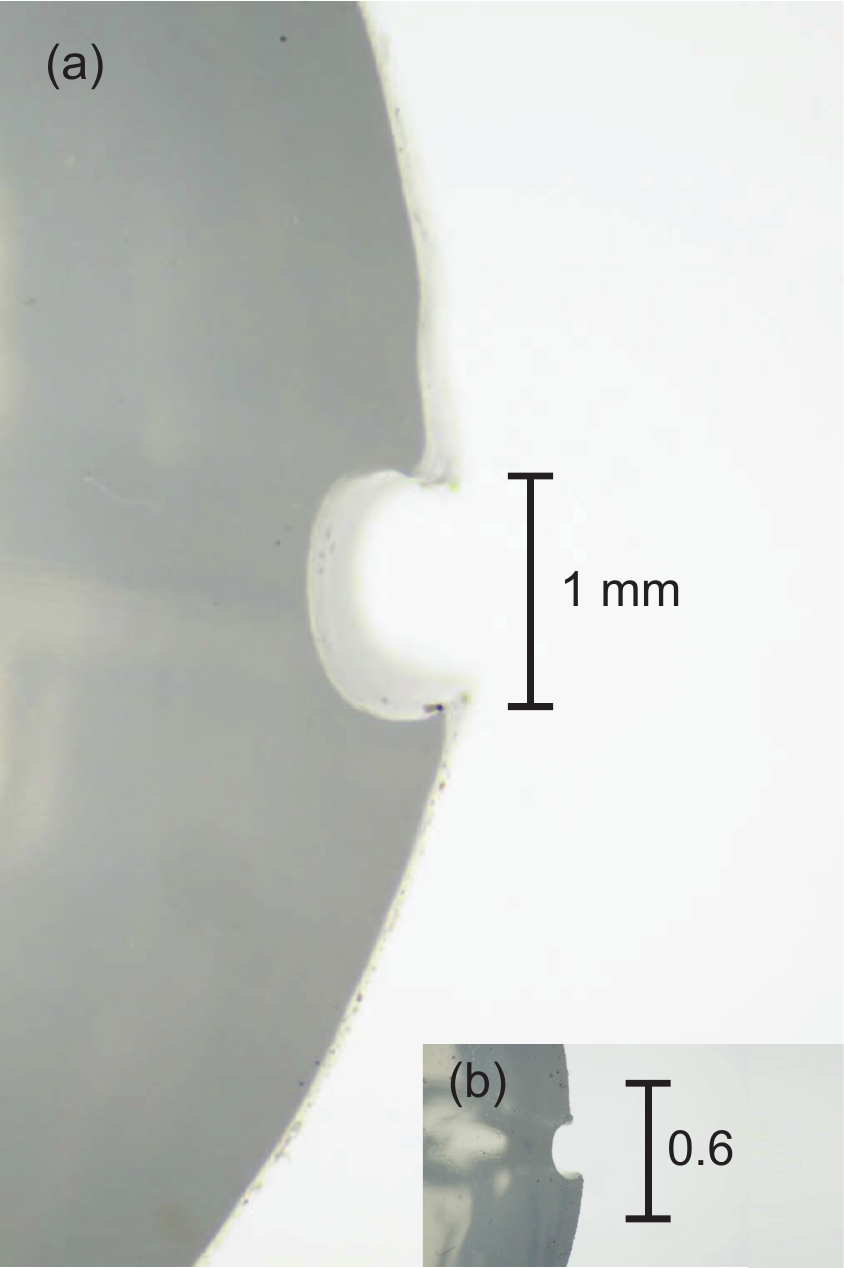}
		\caption{Caption. \label{fig:figBackrind}}
	\end{figure}

	Figure \ref{fig:figBackrind}a is a photograph of SVSR, compression molded in a spherical pocket 1.2 cm in diameter. The finished VSR sphere was cut in half to expose the backrind groove at the parting line. That groove is approximately cylindrical in shape and has a pronounced undercut at its opening. These features are consistent with the groove surface being an isochron as defined above.

	Figure \ref{fig:figBackrind}b shows the backrind that formed when the volume of cured SVSR extruded per unit length was substantially reduced. This backrind is again a cylindrical groove with a pronounced undercut at its opening. Its surface is another isochron, but with an enclosed volume per unit length that is about 40\% that enclosed by the (a) groove surface isochron.

	When curing was complete and the mold was opened, each SVSR sphere was complete and had only a thin flash at its parting line. No groove was visible. However, once the flash was removed and the VSR sphere was allowed to adopt its equilibrium shape, the groove gradually appeared. Over approximately a one-hour period, SVSR at the seam line retracted into the sphere to form the grooves shown in Fig. \ref{fig:figBackrind}. In effect, that SVSR retraced its flow path and returned to where it was when it cured, before thermal expansion pushed the mold open.
	
	\section{Conclusion}
	
	When the dynamic viscoelastic stresses that a cured (solid) thermoset polymer experiences in a situation overwhelm the static elastic stresses, the polymer flows like a liquid. Such liquid-like flow can be observed during the molding of soft viscoelastic silicone rubbers. When rising pressure in a closed mold pushes that mold open and cured SVSR near the parting lines is propelled toward and through gaps at the part lines, it behaves like liquid and undergoes dissipative laminar flow.
	
	Cured SVSR that does not enter the parting line gap undergoes only modest shear and elongation and suffers little or no damage to its network of permanent crosslinks. It thus retains its equilibrium shape. Cured SVSR that is extruded through the parting-line gap, however, undergoes severe shear and elongation and its network of permanent crosslinks is destroyed. It loses its equilibrium shape.
	
	The result of this liquid-like flow and partial-loss of equilibrium shape is a backrind groove. The skin of that groove consists of cured SVSR that barely avoided entering the parting-line gap and thus retained its equilibrium shape. It subsequently retracted back inward as the groove formed. The innermost portion of flash consists of cured SVSR that did enter the gap and lost its equilibrium shape.
	
	In thermosets that experience stronger static elastic stresses, those stresses compete with dynamic viscoelastic stresses and liquid-like flow is no longer possible. Cured thermoset may still move toward and through parting line gaps when rising pressure pushes a mold open, but that movement is complicated, as is the shear and elongation. The polymer can be damaged or torn even before it enters the parting-line gap, leading to unpredictable backrind in the finished product. Though not presented here, such messy, erratic backrinds can be observed in firm VSRs, as they can be in other pliable thermoset rubbers. 

 \bibliography{backrind}
 
\end{document}